\documentclass[conference]{IEEEtran}
\ifCLASSOPTIONcompsoc
  \usepackage[nocompress]{cite}
\else
  \usepackage{cite}
\fi
\ifCLASSINFOpdf
  \usepackage[pdftex]{graphicx}
  \graphicspath{{.}{fig/}}
  \DeclareGraphicsExtensions{.pdf,.jpg,.jpeg,.png,.pdf}
\else
\fi
\usepackage{amsmath}
\usepackage{amsfonts}

\usepackage{tabularx}
\usepackage{multirow}
\usepackage{listings}

\begin{document}
%
\title{\Large \bf 
The usability canary in the security coal mine:\\
A cognitive framework for evaluation and design of usable authentication solutions}



%
\author{\IEEEauthorblockN{Brian Glass\IEEEauthorrefmark{1},
Graeme Jenkinson\IEEEauthorrefmark{2},
Yuqi Liu\IEEEauthorrefmark{1}, 
M. Angela Sasse\IEEEauthorrefmark{1}, 
and Frank Stajano\IEEEauthorrefmark{2}}
\IEEEauthorblockA{\IEEEauthorrefmark{1}University College London}
\IEEEauthorblockA{\IEEEauthorrefmark{2}University of Cambridge}}


\IEEEoverridecommandlockouts
\makeatletter\def\@IEEEpubidpullup{9\baselineskip}\makeatother
\IEEEpubid{\parbox{\columnwidth}{Permission to freely reproduce all or part
    of this paper for noncommercial purposes is granted provided that
    copies bear this notice and the full citation on the first
    page. Reproduction for commercial purposes is strictly prohibited
    without the prior written consent of the Internet Society, the
    first-named author (for reproduction of an entire paper only), and
    the author's employer if the paper was prepared within the scope
    of employment.  \\
    EuroUSEC '16, 18 July 2016, Darmstadt, Germany\\
    Copyright 2016 Internet Society, ISBN 1-891562-45-2\\
    http://dx.doi.org/10.14722/eurousec.2016.23007
}
\hspace{\columnsep}\makebox[\columnwidth]{}}

\maketitle 
 
\begin{abstract} Over the past 15 years, researchers have identified an
increasing number of security mechanisms that are so unusable that the intended
users either circumvent them or give up on a service rather than suffer the
security.  With hindsight, the reasons can be identified easily enough: either
the security task itself is too cumbersome and/or time-consuming, or it creates
high friction with the users' primary task.  The aim of the research presented
here is to equip designers who select and implement security mechanisms with a
method for identifying the ``best fit'' security mechanism at the design stage.
Since many usability problems have been identified with authentication, we
focus on ``best fit'' authentication, and present a framework that allows
security designers not only to model the workload associated with a particular
authentication method, but more importantly to model it in the context of the
user's primary task.  We draw on results from cognitive psychology to create a
method that allows a designer to understand the impact of a particular
authentication method on user productivity and satisfaction. In a validation
study using a physical mockup of an airline check-in kiosk, we demonstrate that
the model can predict user performance and satisfaction. Furthermore, design
experts suggested personalized order recommendations which
were similar to our model's predictions. Our model is the first that supports
identification of a holistic fit between the task of user authentication and
the context in which it is performed. When applied to new systems, we believe
it will help designers understand the usability impact of their security
choices and thus develop solutions that maximize both.
\end{abstract} 
 

%
\IEEEpeerreviewmaketitle

\section{Introduction} 
 
Over the past 15 years, the security community has started to acknowledge that security mechanisms are only effective if they are usable: users frustrated by overzealous security measures bypass the security if they can, or switch to a competing system that is easier to use.  While an increased awareness of the damage that lack of usability can inflict is a first step, in practice security experts and developers who choose security mechanisms have no way of gauging what the impact of their choice on users will be---and most are not able to call on a human usability expert to do this for them.  There are tools for developers to carry out walkthroughs and assessments of a particular solution. The time it will take a user to complete a task can be estimated using the Keystroke Level Modelling (KLM-GOMS) model  \cite{John94thegoms}, and an automated version  CogTools \cite{john2012tools} provides such a prediction from screen interaction with the tool. This approach, however, has limitations: 
 \begin{enumerate} 
 
\item It only supports evaluation and comparison of specified solutions, rather than discovery of the ``best'' one, and 
    
\item it does not take account of the impact that different mental and physical tasks have on subsequent tasks. 
\end{enumerate} 

In this paper, we contribute and validate an intellectual tool---a design and evaluation framework---that will help designers gain a better understanding of the cost of security, with specific reference to user authentication. Our framework and methodology assesses security mechanisms not in isolation but in the context of the so-called \emph{primary task} that constitutes the user's true goal.  What users really want (primary task) is to check in for a flight or pay a bill, not recall and enter a password or read off and transcribe a one-time code. From the users' perspective, these are distractions imposed in the name of security, often to manage threats they don't know exist.  
 
The cost of a given security measure, such as entering a password, is not absolute: it is instead also a function of its relationship to the other components of the primary task.  A recent study \cite{sasse2014great} found that authentication creates a ``wall of disruption'' in users' work. This is not only the time spent on the security task, but the knock-on effect of re-starting the primary task after an interruption.  Thus, the cost depends not just on how hard the authentication task is in itself but also on when it occurs in the user’s workflow, on what functions of the brain it loads and on what else the user was meant to be doing before and after.  

We draw on results from cognitive psychology to assess the cost of task switching between different activities.  Our framework lets designers model the tasks of the intended scenario and the precedence constraints that describe their relationships, and then quantitatively compares alternatives to suggest combinations that minimize the cognitive load and usability cost to the user\footnote{When our tool, the canary, indicates that the environment has become toxic for the systems users you know it's time to beat a hasty retreat.}. 

In addition to providing this novel methodology, we present a validation study which verifies the tool's insights. Using a physical mockup, we test the tool's optimal ("best") suggestion against its pessimal ("worst") suggestion. Moreover, we surveyed a group of professional designers to test our tool's automatic suggestions against the intution of human experts.
 
\section{Modelling a business process} 
\label{sec:modelling-business-process} 
 
 
A business process (or workflow) is a collection of interrelated tasks that are performed by users in order to achieve some objective. It is often the case that only authorized users may perform certain tasks: in such cases the business process will include one or more tasks requiring explicit user authentication.  Tasks that require authentication impose ordering constraints on the business process (users shouldn't be able to complete a task requiring authorization until they have been authenticated). More generally, the business process may have some freedom in the order in which tasks are performed, that is, the tasks have a partial order.  In such cases, system designers have flexibility to rearrange tasks to maximise the system's usability. 
 
Our goal in modelling a business process is twofold. Firstly, we wish to determine the optimal ordering of the tasks, taking into account the switching costs described in sections \ref{sec:cog-res-trans} and \ref{sec:task-prop-trans}. Secondly, we wish to explore the impact of equivalent but alternative tasks for user authentication.  Thus, our model of a business process must include: 

\begin{itemize} 
\item A representation of the set of steps to be performed, 
\item A set of tasks that can be performed at each step, 
\item Hard constraints that enforce the partial ordering of the tasks, and 
\item Soft constraints that capture the costs of switching between tasks. 
\end{itemize} 
 
 
\subsection{Example: airport check-in kiosk} 
 
Throughout this paper our example will be airport check-in using a self-service kiosk. We are not modelling the kiosk of any particular airline or airport but an imaginary one that combines features we have observed on a variety of real kiosks.  We use this business process as our example because its tasks, listed in Table~\ref{tab:checkin-tasks}, use a range of different cognitive resources, detailed in Table~\ref{tab:checkin-tasks2}. We include cognitive tasks such as making decisions or selections and carrying out checks, as well as physical tasks like attaching luggage tags. The check-in procedure necessarily also includes some form of authentication, but there are multiple ways of achieving that. Helping a designer select the most appropriate authentication mechanism for a specific business process is one of the goals of our framework.  
 
\begin{table*} 
  \centering 
  \begin{tabularx}{\textwidth}{ | l | c | p{35mm} | X | } 
 
    \hline \textbf{Task} & \textbf{Code} & \textbf{Prerequisites} & 
    \textbf{Description} \\ \hline 
 
 
    Select language & LANG & & User selects their preferred language from the 
    displayed options. \\ \hline 
 
    Select airline & AIRL & LANG & User selects their airline from the 
    displayed options. \\ \hline 
 
    Booking reference & BKRF & LANG, AIRL & User enters their booking reference 
    using an touchscreen QWERTY keyboard. \\ \hline 
 
    \textit{Authenticate} & \textit{AUTH} & & User authenticates their 
    identity. \\ \hline 
 
    ~$\triangleright$ Passport scan & AUPS & LANG & User authenticates by 
    scanning the photo page of their passport. \\ \hline 
 
    ~$\triangleright$ Passport information & AUPI & LANG & User authenticates 
    by manually entering their passport information. \\ \hline 
 
    ~$\triangleright$ Insert payment card & AUCC & LANG & User authenticates by 
    inserting their payment card. \\ \hline 
 
    ~$\triangleright$ Password & AUPW & LANG & User authenticates by 
    typing a password (assuming the user has an account with the 
    airline).\\ \hline 
 
    Check forbidden items & FRBN & LANG & User presses a button to confirm that 
    their luggage doesn't contain any of the displayed items. \\ \hline 
 
    Check liquids & LIQH & LANG & User presses a button to confirm that their 
    hand luggage doesn't contain any containers of liquid above a certain 
    volume. \\ \hline 
 
    Check luggage size & DIMH & LANG, AIRL & User presses a button to confirm 
    that their hand luggage is below a certain size. \\ \hline 
 
    Select outbound seat & STSO & LANG, BKRF & User selects their 
    outbound seat by clicking on a plan of the available seats in the 
    airplane. \\ \hline 
 
    Select return seat & STSR & LANG, BKRF & User selects their return 
    seat by clicking on a plan of the available seats in the 
    airplane. \\ \hline 
     
    Buy extra bag & EXBG & LANG, BKRF & User optionally pays for 
    additional luggage by clicking a button and swiping a credit 
    card. \\ \hline 
 
    Confirm & CFRM & LANG, BKRF, \textit{AUTH}, LIQH, DIMH, EXBG & 
    User confirms the details entered so far by reading some text  
    and pressing a button. \\ 
    \hline 
 
    Print luggage tag & PRLT & LANG, EXBG, CFRM & User takes a luggage tag from 
    the machine and attaches it to their luggage. \\ \hline 
    Print boarding pass & PRBP & LANG, CFRM & User takes a boarding pass from 
    the machine. \\ \hline 
 
  \end{tabularx} 
    \caption{\label{tab:checkin-tasks}Airport self-service check-in tasks.} 
 
\end{table*} 
 
\begin{figure} 
  \centering 
  \includegraphics[width=\linewidth]{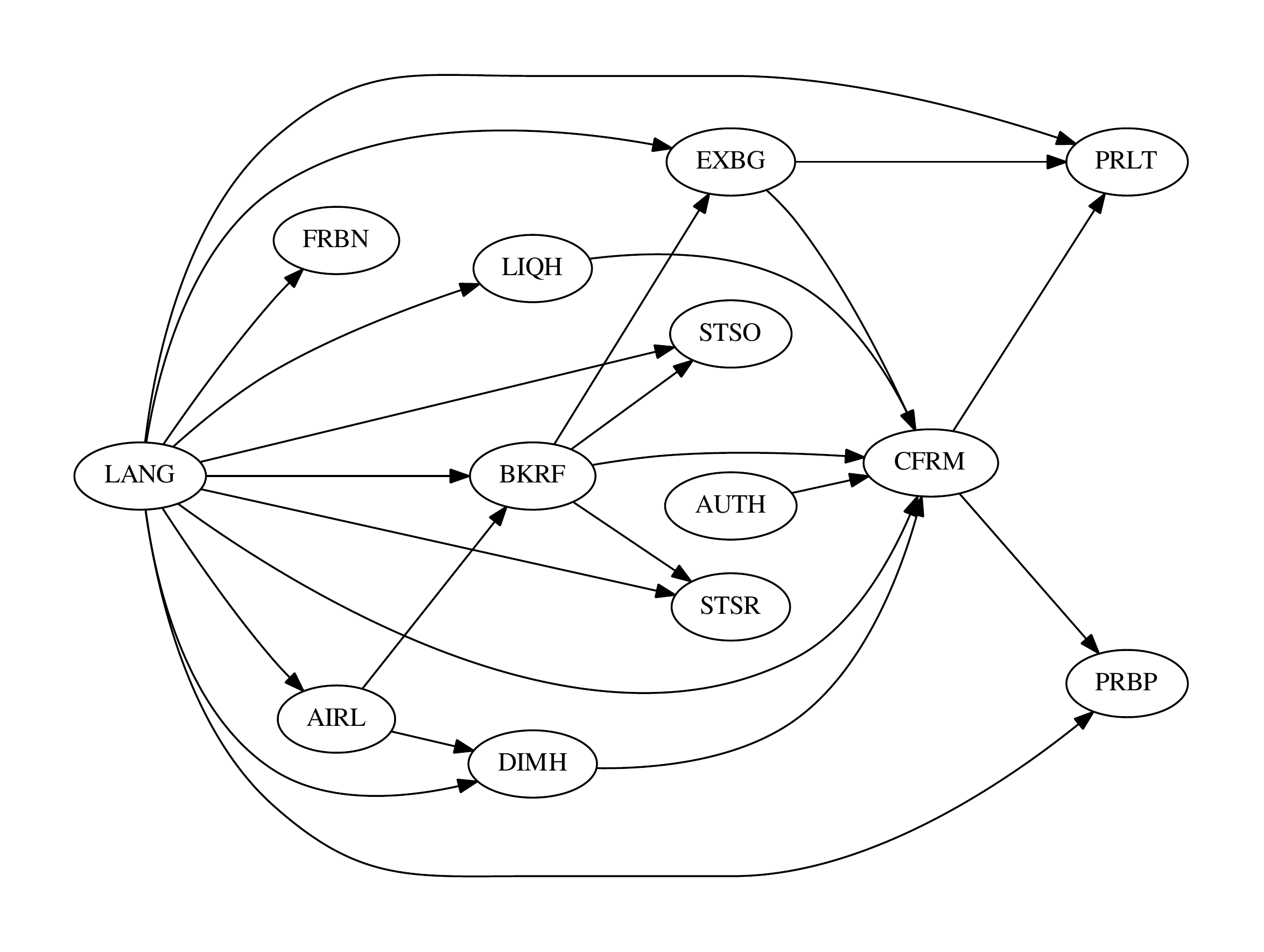} 
  \caption{Dependencies between Airport self-service check-in tasks. An edge from node $u$ to node $v$ indicates that $u$ must be carried out before $v$. For example, users must enter their booking reference (BKRF) sometime before they confirm their check-in (CFRM).} 
  \label{fig:checkin-dag} 
\end{figure}

We are also interested in finding the optimal order for the tasks. The check-in kiosk example exhibits a reasonable degree of ordering flexibility. Figure~\ref{fig:checkin-dag} shows the dependencies between the check-in tasks. 
 
\section{Our framework} 
\label{sec:our-framework} 
 
The framework we present allows developers to assess the usability of different security tasks within a workflow such as the check-in example described above. Two overarching principles inspired this framework: 
 
\begin{enumerate} 
   \item Assessing the usability of an individual task is important, but insufficient, and 
 
   \item the order in which tasks appear can have an interactive, global effect on overall usability. 
\end{enumerate} 
 
Specifically, these principles imply that swapping out one authentication method for another may have carryover effects on the overall workflow. They also suggest that an automated optimization procedure could be used to ``solve'' for the optimal ordering of tasks---the one that minimises cognitive interruptions and maximizes usability. 
 
A workflow has some number of ``steps'' and the user can carry out exactly one task at each step. We want to find the optimal assignment of tasks to steps, respecting any ordering constraints between the tasks---ensuring, for example, that certain tasks happen after the authentication task. We will present a method for encoding a workflow as a weighted constraint satisfaction problem (WCSP) \cite{schiex1995}, in which there are a set of variables (the steps), a set of values (the tasks) and a set of constraints. Further information about constraint satisfaction problems is given in section~\ref{sec:csps} and the means of encoding a workflow is explained in section~\ref{sec:our-model}.  
Workflow environments that are designed to accomplish a specific goal (\textit{e.g.}, withdrawing cash from an ATM, or checking in at an airport kiosk) can be conceptualized as a sequence of tasks completed in a linear fashion.  Transitioning from one task to 
another will carry an additional transition cost. The total usability of an overall task is thus a combination of the costs of the individual tasks and the costs of the pairwise transitions between the tasks in the linearized sequence.  Task ordering can have a potentially unpredictable impact on the entire task workflow.  Considering the usability of many different potential orderings is a non-trivial task but a computerized tool that computes optimal ordering solutions makes it tractable. 
 
The concept that reordering tasks can have an effect on usability comes from established principles in cognitive psychology. An established literature exists on the relative ordering effects of different types of tasks \cite{kiesel2010control}. In section~\ref{sec:cog-psych} we explain how these effects were operationalized from available literature. In this section, we give an overview of how to extend existing assessments of workload by considering not only the endogenous task demands but also the additional exogenous demands that emerge from the transitional costs between tasks. 
 
\subsection{Completion times}
Various methods have been devised to predict the time required to complete a given task. A popular technique is KLM-GOMS \cite{John94thegoms}. In this technique, the designer breaks down the task into a variety of individual action components (for example: mentally prepare, click button, press a key), each of which has an associated reaction time. This technique is useful for estimating how long it would take a user to complete a given task. Another assessment technique is CogTool \cite{john2012tools}, which assesses task completion times and learning rates based on shifting visual attention and making motor responses. Both methods use approximations for mental processes (\emph{think} in CogTool, \emph{mental   preparation} in KLM-GOMS). In the present paper, we seek to expand on these techniques by assessing the differential cognitive demands of different tasks as well as task transitions. 
 
\subsection{Cognitive demands of tasks}
 
While subjective measures of workload are useful tools in predicting user satisfaction and adoption rates, the operationalization of workload as a unitary resource does not fit with modern theories of cognition \cite{deWaard2014,deWinter2014}. Rather, a variety of dissociable mechanisms underlie cognition and become active given characteristics of the task at hand \cite{ashby2011}. In section \ref{sec:oper-checkin-task} we address the various cognitive mechanisms involved in an individual task. 
 
\subsection{Cognitive demands of transitions} 
 
While tasks carry their own demands, there are also certain performance costs associated with switching from one task to another. These transitional costs can be asymmetric; that is, switching from Task A to Task B may be more costly than switching from Task B to Task A \cite{kiesel2010control}. For this reason, we have coded principles of task switching costs from existing literature. In section~\ref{sec:cog-res-trans} and section~\ref{sec:task-prop-trans} we address the various types of switch costs used in the present modelling procedure. 
 
 
 
\subsection{Quantifying tasks} 
 
The goal of our work is to promote a discipline for considering both the unary and transitional demands of tasks on users, and to demonstrate a method for improving performance by minimizing overall task demand. The effectiveness of any given instantiation of this methodology depends directly on the quality of input information about the workflow being analysed. Thus, it will be crucial to develop a valid and reliable regimen for quantifying task characteristics. In this initial paper we are charting a new path and, for illustrative purposes, we have assigned numerical values based on our judgement. In future work we would develop instruments such as worksheets and flowcharts to help independent designers assign consistent and reproducible numerical values when they assess their tasks. 
 
\section{Cognitive psychology} 
\label{sec:cog-psych} 
 
\subsection{Task switching} 
\label{sec:task-switching} 
 
When a person switches from one task to another task, the brain must reorganize and reallocate cognitive resources to ensure an efficient transition \cite{kiesel2010control}. Transitioning from a task that primarily uses resource $A$ to a task that primarily uses resource $B$ (instead of continuing to use resource $A$) results in performance deficits, or switch costs. Experimental psychology has uncovered certain principles that govern these transitions. These so-called \emph{switch cost asymmetries} have been shown to occur, or not, depending on other characteristics of the tasks involved. We have codified these task asymmetries (expressed as Cohen’s $d$ effect sizes, which are a commonly used metric in psychology for comparing the mean of one sample to that of another \cite{ferguson2009effect}) into a collection of rules that may be encoded as constraints in a weighted constraint satisfaction problem (see section \ref{sec:csps}). Below, we describe how we constructed these rules from available literature on switch cost asymmetries. The rules fall into two categories: \emph{cognitive resource transitions} and \emph{task property transitions}. 
 
\subsection{Cognitive resource transitions} 
\label{sec:cog-res-trans} 
 
One reason that task switching results in a performance deficit is the requirement for the individual to disengage active cognitive mechanisms and then engage other cognitive mechanisms in order to match task demands \cite{kiesel2010control}. For example, switching from a visual task to an auditory task is more costly than vice versa \cite{strobach2012task}.  In a practical example, if a person were performing a hypothetical two-factor authentication procedure that involved recognizing an image among several on a large screen and also recognizing a voice over a phone line, it could be more efficient to place the audio identification subtask before the visual authentication subtask. This demonstrates that task ordering can impact user efficiency due to asymmetries in cognitive switch costs.  

\subsubsection{Cognitive resources demands of individual subtasks} 

\label{sec:cog-demands-indiv} 
 
The cognitive mechanisms included in the present implementation are visual working memory (VWM; responsible for holding, processing, and operating on information of immediate importance), procedural memory (PM; responsible for storing and preparing motor action sequences), declarative recall (DR; responsible for generating and presenting stored information on demand), semantic recognition (SR; responsible for determining whether factual information has been stored in memory), and episodic recognition (ER, responsible for determining whether information about experienced events have been stored in memory). Note that while the categories represented here have an empirical basis, the taxonomy of mental processes is a fluid research topic \cite{baddeley2012working}. 
 
Table~\ref{tab:res-trans} reports the costs of switching between tasks utilising different cognitive mechanisms. The values are Cohen`s $d$ effect sizes and were calculated from published studies \cite{ferguson2009effect} involving empirical measurements of reaction time in various task switch contexts, which assessed the efficiency with which individuals were able to transition between different cognitive systems. 
 
\begin{table*} 
  \centering 
 
  \begin{tabular}{ c r|c c c c c| } 
    \cline{3-7} 
    & & \multicolumn{5}{ c| }{\textbf{To}} \\ 
    \cline{3-7} 
    & & VWM & PWM & DR & SR & ER \\ 
    \hline 
 
    \multicolumn{1}{ |c }{\multirow{5}{*}{\rotatebox{90}{\textbf{From}}}} &  
    \multicolumn{1}{ |c| }{Visual working memory (VWM)} 
    & 0 & 0.495 & 0.495 & 0.495 & 0.157 \\ 
 
    \multicolumn{1}{ |c }{} & 
    \multicolumn{1}{ |c| }{Procedural memory (PM)} 
    & 0.495 & 0 & 0.495 & 0.699 & 0.699 \\ 
 
    \multicolumn{1}{ |c }{} & 
    \multicolumn{1}{ |c| }{Declarative recall (DR)} 
    & 0.495 & 0.495 & 0 & 0.482 & 0.482 \\ 
 
    \multicolumn{1}{ |c }{} & 
    \multicolumn{1}{ |c| }{Semantic recognition (SR)} 
    & 0.495 & 0.842 & 1.078 & 0 & 0.433 \\ 
 
    \multicolumn{1}{ |c }{} & 
    \multicolumn{1}{ |c| }{Episodic recognition (ER)} 
    & 0.307 & 0.842 & 1.078 & 0.354 & 0 \\ 
    \hline 
  \end{tabular} 
 
  \caption{\label{tab:res-trans}Costs of switching between tasks  
    utilising different cognitive 
    mechanisms, given as Cohen’s $d$ effect sizes.}   
\end{table*} 
 
\subsubsection{Operationalizing the check-in task} 
\label{sec:oper-checkin-task} 
 
In order to utilise these principles of task switch cost asymmetry, we operationalised identified the cognitive resources most likely to be engaged by the subtasks involved in the Airline Self-Service Task. While this is a first approximation, in the future empirical methods could be used to verify these predictions. In real-world tasks, many different cognitive mechanisms are likely to be engaged simultaneously. For our purposes, we have selected the dominant resource which is predicted to have the highest relative engagement level. Table~\ref{tab:checkin-tasks2} reports the major cognitive resource assigned to each subtask, as well as the physical response modality, voluntary/involuntary nature, task familiarity, and task complexity. 
 
It is impractical to determine the specific brain networks activated for a specific real task, so we characterize each task by assessing its similarity to documented cognitive tasks. For example, determining whether a piece of hand luggage exceeds certain dimensions is similar to documented tasks involving assessing geometric attributes of three dimensional shapes, a task known to activate visual working memory \cite{downing2000}. This is a tractable simplification of the reality of cognitive functioning for two reasons:  
\begin{enumerate} 
       
   \item Real-world tasks likely engage many different cognitive mechanisms at once, with varying degrees of demand. For our purposes we consider the cognitive mechanisms deemed to be most relied upon in order to complete the task. 
 
   \item Many other cognitive mechanisms exist than were included in Table~\ref{tab:res-trans}. For simplicity, we only included the primary mechanisms involved for each task. Future implementations could include other systems such as auditory working memory. 
\end{enumerate}

\subsection{Task property transitions} 
\label{sec:task-prop-trans} 
 An important source of task switch costs is the impact of the interference or inertia carried over from one to another. One counterintuitive finding is that switching from a less familiar task to a more familiar task is actually more disruptive than vice versa \cite{yeung2003switching}. The prevailing reasoning behind this effect is that when engaged in a less familiar task, the individual must suppress commonly used mental processes in lieu of less frequently used processes \cite{gade2007influence}. This suppression has a carry-over effect on the new task, resulting in a performance deficit. These transitional asymmetries have also been identified when transitioning between tasks that differ by complexity \cite{rubinstein2001executive}, recent practice \cite{yeung2003switching}, modality (form or method of response) \cite{sandhu2013modality}, and whether the task was voluntary \cite{arrington2005voluntary}. These empirical observations have been codified into conditional rules with associated effect sizes in Table~\ref{tab:prop-trans}. 
 
\begin{table*} 
 
  \centering 
  \begin{tabularx}{\textwidth}{ | l | X | c | } 
    \hline 
    \textbf{Rule name} & \textbf{Condition} & \textbf{Cost (effect size)} \\ 
    \hline 
 
 
    Modality & A switch occurred which uses the same resources (on-diagonal 
    above) \textbf{and} there is a modality switch. & 0.16 \\ \hline 
 
    Recent Practice & A task of similar modality or resource has been used 
    anywhere previously. & 0.31 \\ \hline 
 
    Familiarity & The current task is more familiar than the previous 
    task. & 0.42 \\ \hline 
 
    Complexity/Choice & A task is done voluntarily \textbf{and} the 
    complexity decreases. & 2.92 \\ 
     
    & A task is involuntary \textbf{and} the complexity decreases. & 1.63 
    \\ \hline 
  \end{tabularx} 
  \caption{\label{tab:prop-trans}Additional costs of transitioning between  
    tasks determined by 
    specific rules, given as Cohen’s $d$ effect sizes.} 
   
\end{table*} 
 
\subsubsection{Complexity} 
 
Task complexity was assessed using existing definitions from experimental psychology \cite{rubinstein2001executive}, namely the number and combination of rules required to solve or complete the task. For example, subtraction is relatively less complex than division. The reason for this is that division uses the principles of subtraction as well as other principles, such as remainders and carrying digits between places. In the airline check-in task, for example, the task regarding forbidden materials was considered to be more complex than the task regarding liquids. This is because it is more complex to determine whether several items fall into several categories versus a single category. 
 
\subsubsection{Familiarity} 
 Task familiarity was determined by assessing not only the frequency with which an average user completes a given task, but also whether the task assesses familiar knowledge or processes \cite{yeung2003switching}. For example, selecting your language preference might not necessarily be a common chore, but it requires judgment based on a familiar fact. In contrast, printing a luggage tag is something that is an activity that is both infrequent and unfamiliar. 
 
\subsubsection{Response Modality} 
 
Response modality refers to the physical method for issuing a response from the user to the system.  For example, different modalities include a QWERTY keyboard, a mouse pointer, or a verbal response. There is evidence that transitioning from one response modality to another can incur a switch cost. However, Sandhu and Dyson \cite{sandhu2013modality} demonstrate that a switch cost due to response modality may not occur when a modality switch coincides with a cognitive resource switch. In other words, switching response modalities is most disruptive when it is the only change that takes place. 
\begin{table*} 
 
  \centering 
  \begin{tabularx}{\textwidth}{ | c | l | X | c | c | c | } 
 
    \hline \textbf{Code} & \textbf{Primary cognitive resource} & 
    \textbf{Modality} & \textbf{Voluntary?} & \textbf{Familiarity} & 
    \textbf{Complexity} \\ \hline 
 
         
    LANG & Semantic recognition & Touchscreen & No & 5 & 1 \\ \hline 
 
    AIRL & Episodic recognition & Touchscreen & No & 5 & 1 \\ \hline 
 
    BKRF & Visual working memory & Touchscreen QWERTY & No & 3 & 3 \\ \hline 
 
    AUPS & Procedural memory & Passport scanner & No & 2 & 2 \\ \hline 
 
    AUPI & Procedural memory & Touchscreen QWERTY & No & 2 & 3 \\ \hline 
 
    AUCC & Procedural memory & Credit card reader & No & 3 & 2 \\ \hline 
 
    AUPW & Declarative recall & Touchscreen QWERTY & No & 4 & 3 \\ \hline 
 
    FRBN & Semantic recognition & Touchscreen & No & 2 & 3 \\ \hline 
 
    LIQH & Episodic & Touchscreen & No & 3 & 3 \\ \hline 
 
    DIMH & Visual working memory & Touchscreen & No & 2 & 4 \\ \hline 
 
    STSO & Visual working memory & Touchscreen & Yes & 2 & 4 \\ \hline 
 
    STSR & Visual working memory & Touchscreen & Yes & 2 & 4 \\ \hline 
 
    EXBG & Episodic & Touchscreen & Yes & 2 & 2 \\ \hline 
 
    CFRM & Episodic & Touchscreen & No & 4 & 2 \\ \hline 
 
    PRLT & Procedural memory & Luggage tag & No & 1 & 5 \\ \hline 
 
    PRBP & Episodic & Touchscreen & Yes & 4 & 2 \\ \hline 
  \end{tabularx} 
  \caption{\label{tab:checkin-tasks2}%
    Properties of the check-in kiosk tasks. Familiarity and 
    complexity are on a scale from 1 (low) to 5 (high).}

\end{table*} 
 
\section{Modelling a business process as a constraint satisfaction problem} 
\label{sec:csps} 
 
\subsection{Constraint satisfaction problems} 
 
The goal of a constraint satisfaction problem (CSP) is to assign \emph{values} to a set of \emph{variables} subject to a set of \emph{constraints}. The constraints express \emph{local} restrictions, such as ``these two variables must have different values''.  An evaluation of the CSP is consistent and complete if it includes all variables and does not violate any constraints (efficient algorithms for finding global solutions are given in \cite{kumar1992}). Below we shall describe \emph{weighted} constraint satisfaction problems: these include ``soft'' constraints that may be violated for some cost. We first introduce the classic CSP framework. 
 
\subsubsection{Classic CSP} 
 
A \emph{classic} CSP is defined by a triple $P = (X, D, C)$. $X$ is the set of variables, $\{x_1, ..., x_n\}$. A domain $d_i \in D$ is a set of allowable values for variable $x_i$. A constraint $c \in C$ is a pair $(X_c, R_c)$, where $X_c \subset X$ is the \emph{scope} of the constraint and $R_c$ is a relation over the corresponding set of domains. $R_c$ specifies tuples of simultaneously-allowed values for the variables in the scope and can be defined explicitly as a subset of the product of the domains, or as an abstract relation which can test whether a given tuple of values is allowed, for example: $x_1 \neq x_2$. 
 
An \emph{assignment} specifies values for some or all of the variables. An assignment is \emph{consistent} if it does not violate any constraints. A \emph{complete} assignment is one which assigns values to all variables. A \emph{solution} to a CSP is a complete consistent assignment. A CSP is consistent if a solution for it exists. Finding a solution to a CSP is an NP-complete problem. 
 
\subsubsection{Weighted CSP} 
\label{section:weightedcsp} 
 
In a classical CSP the constraints are all absolute or ``hard'', no consistent assignment can violate any constraint and all solutions are equally ``good''. Several variants have been proposed to extend the CSP framework to include ``soft'' constraints expressing priorities, preferences, costs, and probabilities.  Schiex, Fargier and Verfaillie \cite{schiex1995} generalised these and defined \emph{valued CSP} (VCSP).  A VCSP is similar to a classical CSP except that the constraints assign \emph{costs} to assignments instead of allowing or disallowing them\footnote{Equivalently a VCSP can be seen as   classic CSP in which each constraint has been annotated with a cost for   removing it \cite{schiex1995}.}. 
 
A VCSP is defined by a tuple $P = (S, X, D, C)$, where $X$ and $D$ are sets of variables and their domains as previously.  Costs are specified using a \emph{valuation structure}, which is a triple $S=(E, \oplus, \succ)$, where $E$ is a set of costs ordered by $\succ$ and $\oplus$ is an associative commutative monotonic binary operation on $E$ for combining costs.\footnote{A classical CSP   can be expressed as a VCSP with $E = \{t, f\}$, $\bot=t \succ f=\top$ and   $\oplus = \wedge$.}  Weighted CSP (WCSP) is a specific subclass of valued CSP in which the costs are the natural numbers or positive infinity, $E = \mathbb{N} \cup \{\infty\}$ and $\oplus$ is the standard sum operation. 
 
In this framework, constraints specify local costs of assignments.  A constraint $c \in C$ is a pair $(X_c, F_c)$ where $X_c$ is its scope and $F_c$ is a cost function,

\begin{equation} 
  F_c : \prod_{x_i \in X_c} d_i \to E 
\end{equation} 
 
Note that a hard CSP constraint $c = (X_c, R_c)$ can be 
represented in a WCSP as $c' = (X_c, F_{c'})$, where 
 
\begin{equation} 
  F_{c'}(v) = \left\{ 
  \begin{array}{ll} 
    0 & \text{if } v \in R_c\\ 
    \infty & \text{otherwise} 
  \end{array} 
  \right. 
\end{equation} 
 
Given a WCSP $P = (S, X, D, C)$, an assignment $A$ of variables $Y \subset X$ has total cost $V_P(A) \in E$. This cost is the sum of all applicable cost functions. 
 
 
%
 
Given a WCSP, the typical task is to find the optimal solution, the complete assignment with the minimum total cost. The most popular algorithms for solving WCSP employ branch and bound search, although algorithms for solving WSCP remain an active research area. 
 
 
\subsection{Our model} 
\label{sec:our-model} 
 
As described in section \ref{section:weightedcsp}, a weighted CSP is represented by the tuple $P = (S, X, D, C)$.  In our model, a business process with $n$ steps (where $1$ is the first step performed by the user and $n$ the last) is represented by a set of variables $X$, $\{x_1, ..., x_n\}$.  The domain $D$ (the set of values that can be assigned to variable $x_i$) consists of all of the tasks, including any user authentication tasks, in the business process.  The set of constraints $C$ includes hard constraints that ensure tasks are performed exactly once and ordering relations between tasks are maintained.  $C$ also includes soft constraints represent the costs of switching between tasks. 
 
\subsubsection{Implementation of our model} 
 
A proof-of-concept implementation of our model has been created in Numberjack, a Python framework for constraint programming, mixed integer programming and satisifiability solvers \cite{hebrard10:con}. Numberjack integrates a number of third-party, open source solvers (which are typically written in C/C++ for efficiency) and can be easily extended to include additional solvers.  The Numberjack framework includes support for Toulbar2---an exact combinatorial optimization tool designed for solving Weighted Constraint Satisfaction Problems (otherwise known as Cost Function Networks) \cite{allouche2010toulbar2}. Numberjack's proposition of a high-level modelling framework and an underlying efficient and high-pedigree solver\footnote{Toulbar2 was a wining solver in the Uncertainty in   Artificial Intelligence (UAI) 2010 Approximate Inference Challenge.} make it well suited to our purpose. 
 
As shown below, a Numberjack \texttt{VarArray} is used to represent each step in the business process. The domain of each variable is the natural numbers $0...d$ where each value represents one of the possible tasks. A constraint is then added to the model to ensure that each value in the domain is assigned to exactly one variable. 
 
\begin{lstlisting}[ 
   language=Python, 
   basicstyle=\ttfamily\small, 
   keywordstyle=\ttfamily\small, 
   breaklines=true, 
   breakatwhitespace=true] 
from Numberjack import VarArray 
 
# Create a variable array, 
# one variable for each step  
# in the business process 
wcspVars = VarArray(0, d, nSteps) 
 
model.add(AllDiff(wcspVariables)) 
\end{lstlisting} 
 
A custom Numberjack constraint has been created to enforce the partial ordering of tasks. This constraint (shown below) ensures that for all combinations of the variables in the CSP it is never the case that the value \textit{after} is assigned to a variable that precedes a variable assigned the value \textit{before}. 
 
\begin{lstlisting}[ 
   language=Python, 
   basicstyle=\ttfamily\small, 
   keywordstyle=\ttfamily\small, 
   breaklines=true, 
   breakatwhitespace=true] 
 
class Order(Predicate): 
 
    def __init__(self, vars, before, after): 
        Predicate.__init__(self, vars, "Order") 
        self.set_children(vars) 
        self.before = before 
        self.after = after 
        self.lb = None 
        self.ub = None 
 
    def decompose(self): 
        return [(x != self.after) | (y != self.before) for x, y in combinations(self.children, 2)] 
\end{lstlisting} 
 
As defined in section \ref{section:weightedcsp}, a constraint $c \in C$ is a pair $(X_c, F_c)$ where $X_c$ is its scope and $F_c$ is a cost function. Task switching costs are modelled as binary constraints; that is, their scope is limited to variables that are immediately next to each other. The task switching costs are represented by a $d$-by-$d$ matrix (where $d=|D|$). 
 
\begin{lstlisting}[ 
   language=Python, 
   basicstyle=\ttfamily\small, 
   keywordstyle=\ttfamily\small, 
   breaklines=true, 
   breakatwhitespace=true] 
from Numberjack import PostBinary  
 
def pairwise(iterable): 
    a, b = tee(iterable) 
    next(b, None) 
    return izip(a, b) 
 
# d-by-d matrix, 
# binaryCost[d1][d2] specifies the  
# cost of assigning d1 and d2 to 
# variables that are immediately 
# next to each other 
binaryCosts = [...] 
 
for var, varNext in pairwise(wcspVars): 
   model.add(PostBinary(var, varNext, binaryCosts))  
\end{lstlisting} 
 
\subsubsection{Results of modelling the airline self service check-in} 
 
\begin{table*} 
  \centering 
  \begin{tabularx}{\textwidth}{ lXXXX } 
    \cline{2-5} 
 
    & Select language & Select language & Select language & Select language \\ 
    \cline{2-5} 
 
    & Select airline & Select airline & Select airline & Select airline \\ 
    \cline{2-5} 
     
    & Check liquids & Check liquids & Check liquids & Check liquids \\ 
    \cline{2-5} 
 
    & Booking reference & Booking reference & Booking reference & Booking 
    reference \\ \cline{2-5} 
 
    & Check forbidden items & \textbf{Insert payment card} & 
    \textbf{Passport info} & \textbf{Password} \\ \cline{2-5} 
   
    & Select return seat & Buy extra bag & Select return seat & Check forbidden 
    items \\ \cline{2-5} 
 
    & Check luggage size & Select return seat & Check luggage size & Select 
    outbound seat \\ \cline{2-5} 
 
    & \textbf{Passport scan} & Check luggage size & Check forbidden items & Check luggage 
    size \\ \cline{2-5} 
 
    & Buy extra bag & Check forbidden items & Buy extra bag & Buy extra bag \\ 
    \cline{2-5} 
 
    & Confirm & Confirm & Confirm & Confirm \\ \cline{2-5} 
 
    & Print boarding pass & Print boarding pass & Print boarding pass & 
    Print boarding pass \\ \cline{2-5} 
 
    & Select outbound seat & Select outbound seat & Select outbound seat & 
    Select return seat \\ \cline{2-5} 
 
    & Print luggage tag & Print luggage tag & Print luggage tag & Print luggage 
    tag \\ \hline\hline 
 
    Cost & 5.53 & 5.88 & 8.18 & 8.42 \\ \hline 
 
  \end{tabularx} 
 
  \caption{\label{tab:optimal-order}%
    Optimal task ordering of the self-service check-in using different 
    authentication mechanisms.} 
 
\end{table*} 
 
Table \ref{tab:optimal-order} shows the optimal task ordering given by the solver for the self-service check-in scenario. The four columns of the table correspond to the four different concrete authentication tasks we are considering. The cost reported for each workflow is the sum of all the task switch costs (Cohen's $d$ effect sizes) for that workflow\footnote{To obtain the total cost, we should add to that the   costs of the individual subtasks. We cannot do that yet, because   they are expressed in different non-comparable units, so this is a   topic for future research. See the next section,   \ref{limitations}.}.  The fact that the four orderings and total costs are different supports the central message of this paper: fitting an authentication task to its context is important. Specifically, we can see that the passport scan (AUPS) and insert payment card (AUCC) authentication methods yield substantially lower total switching costs---\emph{regardless} of their intrinsic costs. More generally, with twelve task switches in total, the mean cost for each task switch, in each of the four cases, is approximately 0.5, which constitutes a ``medium'' effect size under the standard Cohen's $d$ interpretation: this indicates that task switches are not an insignificant cost in general. 
 
It is interesting to note that the solver splits the two seat selection tasks for the outgoing and return flight.  Within the model, the two selection tasks are indistinguishable so the cost of switching from either to the other is zero. Therefore, we might expect that the solver would place these task next to each other. However, this is an interesting example of how our intuition can be wrong as this local optimization ultimately precludes the globally optimal solution.  
\subsubsection{Limitations of our model} 
\label{limitations} 
 
\begin{quote} 
  Essentially, all models are wrong, but some are useful. 
  \hfill---\emph{George E. P. Box}~\cite{box1987} 
\end{quote} 
 
The first significant limitation of our model is its inability to relate the reported total task switching costs to an additional amount of \emph{time} required to complete the business process. Whilst this is a significant limitation, we feel that the outputs of the model remain useful and may be used alongside the existing techniques for estimating the time taken to carry out specific tasks such as KLM-GOMS. 
 
Secondly, although the cognitive resource transition costs and task property transition costs are based on empirical results from the literature, user studies should be undertaken to validate the way in which they combine within our framework.  

As well as splitting up the two seat selection tasks, in three cases the solver has placed return seat selection before outbound seat selection. While this would obviously be somewhat confusing for users, it is understandable that the solver has arranged the tasks in this way because within the model they appear identical. Our model simply doesn't capture the notion that when tasks relate to events that are ordered, it makes sense for those tasks to have the same order. In such cases the system designer must apply their discretion to ensure that the system remains consistent with reality and with user expectations. 
 

\begin{figure} 
  \centering 
  \includegraphics[width=\linewidth]{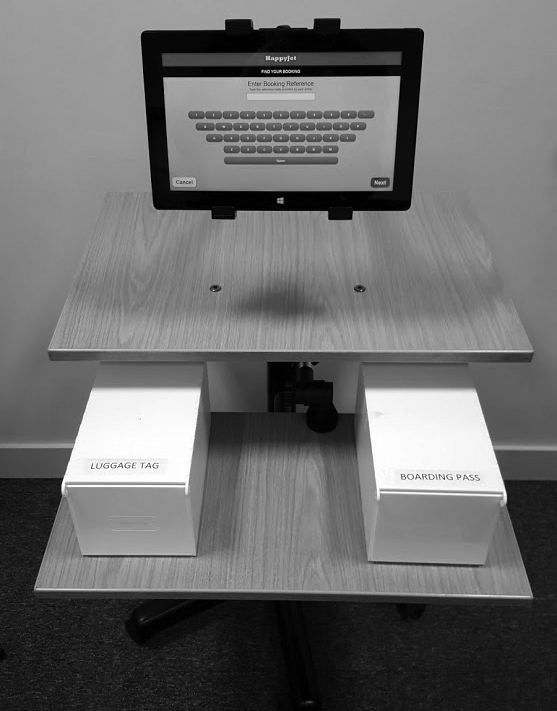} 
  \caption{Mock up for the self-service airport check-in kiosk.} 
  \label{fig:kiosksetup_bw} 
\end{figure}

\section{Validation Study}

In order to test the model's predictions, we completed a validation study. Our intention was to validate the theoretical predictions regarding task switching, and thus we focused on the subtasks which would be inherent in airline check-in kiosks regardless of further authentication mechanisms used (e.g., credit card, passport). Using a mock-up of the airline check-in kiosk described above, we sought to assess the model's optimal subtask ordering recommendation. We accomplished this in four ways: 1) Participants completed the optimal ("best") ordering in a simulated airline departure scenario, 2) These same participants offered their own order recommendations for the task, 3) We further tested a second sample of participants with the pessimal ("worst") ordering, and 4) We surveyed professionals trained in design fields in order to gather an expert based ordering recommendation. The Optimal ordering was: AIRL, LIQH, BKRF, FRBN, STSO, DIMH, EXBG, CFRM, PRBP, PRLT; the Pessimal ordering was: FRBN, AIRL, BKRF, EXBG, LIQH, DIMH, CFRM, PRLT, STSO, PRBP.

\subsection{Participants}

Participants were recruited from the University College London student and staff community and compensated \pounds7 for their time. The study was approved by the UCL Ethics Committee, and all participants offered informed consent. For the Optimal condition, 40 participants were recruited. A sample of 20 participants was recruited for the comparative Pessimal condition, and a further 50 self-reported design professionals were recruited to generate the Expert ordering suggestion. The demographics of the group were as follows: Optimal group ($Age_{Mean} = 26.6$, $Age_{SD} = 7.2$, 28 females), Pessimal group ($Age_{Mean} = 29.1$, $Age_{SD} = 13.5$, 15 females), Expert designers ($Age_{Mean} = 30.0$, $Age_{SD} = 9.7$, 12 females, 8 no gender specified). Two participants were removed from the Optimal group for not completing the task, and three were removed from the Expert group for not completing the survey. Participants were asked about their average annual number of flights: $Optimal Group = 4.7 (SD = 3.4)$, $Pessimal Group = 3.5 (SD = 3.2)$.

The sample of Expert designers was recruited from NCR Corporation (www.ncr.com) as well as via the online survey system Prolific Academic (www.prolific.ac), and were selected using a pre-screening occupation questionnaire. The group identified as working with user experience design in physical settings (n=17), software/web settings (n=33), or both (n=6), with 5.2 mean years of experience ($SD$ = 6.0). These participants were compensated with \pounds1.67 for completing the task (equivalent to \pounds5/hour).

\subsection{Procedure}

\subsubsection{Check-in Kiosk}
Participants were asked to use the simulated airline check-in kiosk as if they were actually preparing for a departure at an airport. Participants were given two suitcases, one large suitcase for checked baggage, and one small suitcase for carry on. The experimenter opened the small suitcase and described the contents to the user: two shirts, two paperback books, and a plastic bag containing toiletries under 100ml in volume. The experimenter told the participant that the large suitcase contained clothes and no hazardous or forbidden materials. The participants completed the airline check-in kiosk three times, each time with a different provided cover story (given in pseudo-random order between participants). The mock airlines were ``MetroAir'', ``HappyJet'', and ``QuickFly'', and the mock destinations were Glasgow, Edinburgh, and Cardiff (departing from London). Participants took the two suitcases and entered a second room to interact with a kiosk comprised of a touchscreen monitor and two dispensers (one for boarding pass, one for baggage tag) on a small roller table (see Figure~\ref{fig:kiosksetup_bw}). The dispensers were pre-loaded with the relevant boarding pass and baggage tag, and a simulated printing sound oriented the participant to their locations during the appropriate subtask. After completing each of the three simulated check-in procedures, the participant moved to a different room and completed the subjective satisfaction questionnaire.

\subsubsection{Subjective Satisfaction Questionnaire}

After each trial, participants completed the following 13-item Satisfaction Questionnaire \cite{comrey1988factor}. Each item was scored using a 5-point Likert scale (from "Strongly disagree" to "Strongly agree"). In order to reduce repetitiveness, the second and third repetitions of the questionnaire asked for changes in assessment relative to the previous trial (from "Less than before" to "More than before"). In this way, a change score was computed using responses from the first trial as a baseline.

\begin{enumerate}
\item The system was annoying to use.
\item I liked using the system.
\item The system did what I thought it would do.
\item The system was fun to use.
\item The system was unreliable.
\item I was satisfied using this system.
\item I was comfortable using this system.
\item The system was disappointing.
\item The system was engaging.
\item The system was unpredictable.
\item I feel positive about the system.
\item I would not want to use this system.
\item The system was pleasant to use.
\end{enumerate}

\subsubsection{Ordering Preference Task}

\begin{figure} 
  \centering 
  \includegraphics[width=\linewidth]{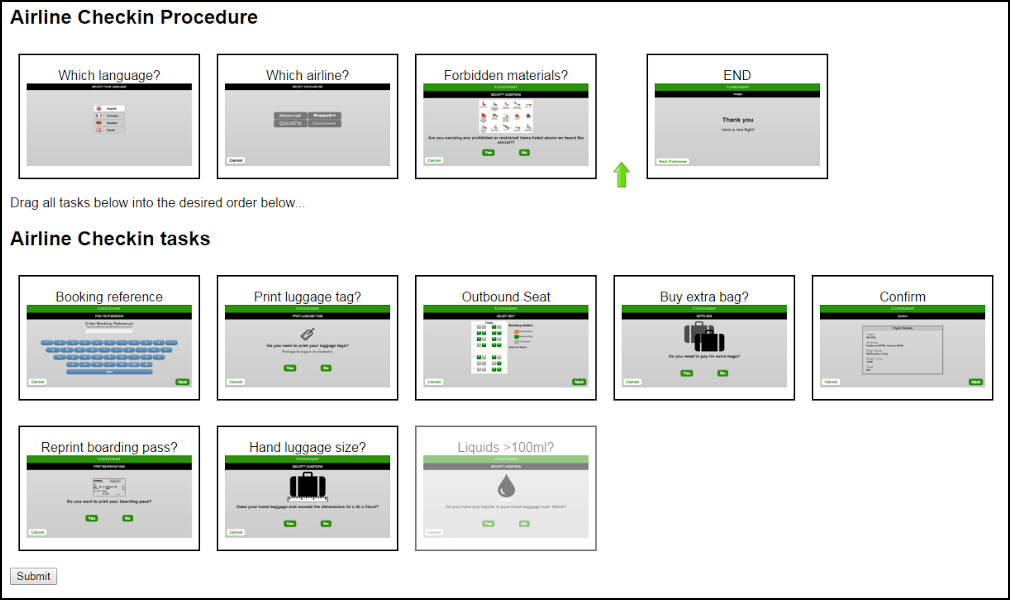} 
  \caption{Screenshot of the ordering preference task.} 
  \label{fig:ordertaskshot} 
\end{figure}

After the completion of the check-in procedure, participants were asked to generate their own suggested orderings for the subtasks. Using a computerized tool, participants dragged boxes representing the various subtasks into their preferred orderings. First, participants were allowed to freely order the subtasks without partial ordering constraints. Second, participants were told which subtasks violated the partial ordering constraints (if any), and were asked to rearrange the subtasks until the ordering satisfied the constraints (see Figure~\ref{fig:ordertaskshot}).

\subsection{Results}

\subsubsection{User Performance}

Task performance was measured by calculating the time to complete each subtask. The time was computed based on the duration from completion of previous subtask to the completion of the current subtask. Results were similar when time was calculated as the duration from the completion of the previous subtask to the first click of the current subtask, although some subtasks only required one click, thus we present subtask completion times here.

\begin{figure} 
  \centering 
  \includegraphics[width=\linewidth]{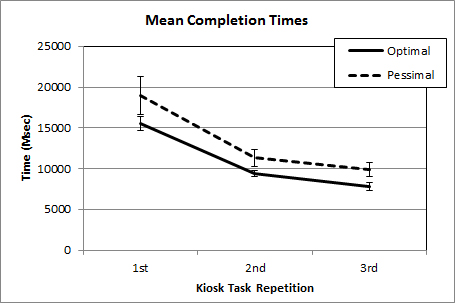} 
  \caption{Mean completion times over three task repetitions between Optimal ordering and Pessimal ordering conditions.} 
  \label{fig:completion_times_mean} 
\end{figure}

\begin{figure} 
  \centering 
  \includegraphics[width=\linewidth]{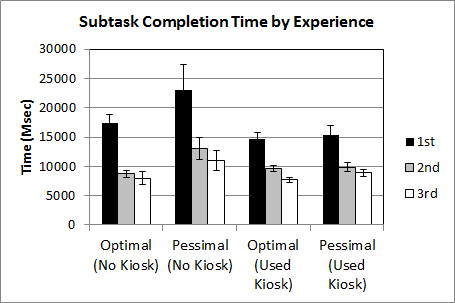} 
  \caption{Mean completion times over three task repetitions between Optimal ordering and Pessimal ordering conditions, by experience.} 
  \label{fig:completion_times_experience_mean} 
\end{figure}

To evaluate the impact of our model's ordering suggestions as well as the impact of prior kiosk experience, participants were further clustered into two experience groups: Have used airline check-in kiosk in the previous calendar year (Used Kiosk), or have not (No Kiosk). Learning curve (repetition over the three trials) was also evaluated as a within subjects factor. Performance (mean completion time) was evaluated using a repeated measures ANOVA with a 2 (Condition: Optimal, Pessimal) x 2 (Experience: Used Kiosk, No Kiosk) x 3 (Repetition) factorial design. There were significant main effects of Condition ($F_{1,55} = 4.82, p = 0.03$) and Experience ($F_{1,55} = 5.01, p = 0.03$) such that those in the Optimal order had faster completion times, and those with airline kiosk experience in the previous year had faster completion times. There was a significant main effect of Repetition ($F_{2,110} = 81.0, p < 0.001$) consistent with a monotonic learning curve (see Figure~\ref{fig:completion_times_mean}). There was also a significant interaction of Repetition and Experience ($F_{2,110} = 5.09, p = 0.01$) such that those with experience demonstrated a flatter learning curve due to faster initial completion times (see Figure~\ref{fig:completion_times_experience_mean}). Completion time was lower for 8 out of 10 subtasks (essentially tied for PRBP and AIRL). According to the binomial distribution, the probability of a result at least this extreme occurring from randomly generated data is 5.3\%. In summary, those in the Optimal ordering condition demonstrated faster completion times on all three repetitions of the task, and those with prior experience were overall faster as well.

\subsubsection{User Satisfaction}

\begin{figure} 
  \centering 
  \includegraphics[width=\linewidth]{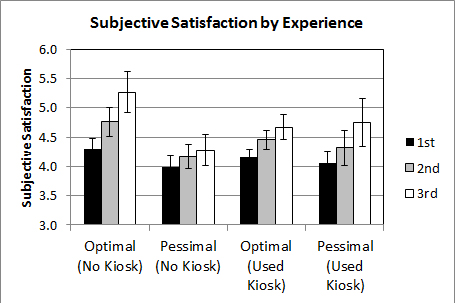} 
  \caption{Satisfaction Scores over three task repetitions between Optimal ordering and Pessimal ordering conditions, by experience.} 
  \label{fig:satisfaction_experience} 
\end{figure}

User satisfaction was measured using the 13-item Satisfaction Questionnaire (see above) by taking the average responses on a 5-point Likert scale (reverse coded for the negatively worded items). For the second and third completion of the questionnaire, the scores were demeaned (subtracted by 3) and added to the previous questionnaire's result. Satisfaction was evaluated using a repeated measures ANOVA with a 2 (Condition: Optimal, Pessimal) x 2 (Experience: Used Kiosk, No Kiosk) x 3 (Repetition) factorial design.  Although directionally in favor of the Optimal ordering, the satisfaction ratings were not statistically significantly higher for the Optimal ordering versus the Pessimal ordering ($F_{1,55} = 2.15, p = 0.149$). There was a significant main effect of Repetition ($F_{2,110} = 27.9, p < 0.001$) such that subjective user satisfaction increased monotonically over the three task repetitions. There was a significant three-way interaction of Repetition, Condition, and Experience ($F_{2,110} = 3.68, p = 0.03$). Figure~\ref{fig:satisfaction_experience} illustrates the nature of this interaction, such that those with no kiosk experience were more sensitive to the Optimal vs. Pessimal manipulation than those with kiosk experience. Specifically, those with no kiosk usage in the previous year found the Optimal ordering to be more satisfactory over time relative to the Pessimal ordering.

\subsubsection{Ordering Preferences}

Participants provided their own recommended orderings for the kiosk subtasks. Separately, self-reported design professionals (who did not complete the kiosk task) also provided recommended orderings. From these experts, a consensus Expert ordering was generated (AIRL, BKRF, STSO, DIMH, FRBN, LIQH, EXBG, CFRM, PRLT, PRBP) using the mode frequencies from each subtask index. A Euclidean distance metric (based on index differences) was computed for each participant's recommended ordering. In this way, we were able to calculate a participant's suggestion's difference from the model's Optimal ordering, Pessimal ordering, and an Expert ordering. The Expert ordering was significantly more similar to the model's Optimal ordering than the Pessimal ordering ($t_{Paired} = 9.15, p < 0.001$). Thus, Experts suggested orderings which were more similar to the model's Optimal suggestion.

\begin{figure} 
  \centering 
  \includegraphics[width=\linewidth]{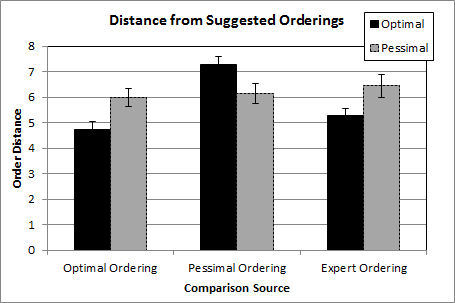} 
  \caption{Difference of participants' suggested orderings to the model's Optimal and Pessimal ordering, as well as an Expert suggested ordering.} 
  \label{fig:suggested_ordering_distance} 
\end{figure}

Prefered ordering was evaluated using a repeated measures ANOVA with a 2 (Condition: Optimal, Pessimal) x 2 (Experience: Used Kiosk, No Kiosk) x 3 (Comparison Source: Optimal, Pessimal, Expert) factorial design. There was a significant main effect of Comparison Source ($F_{2,110} = 27.4, p < 0.001$) and a significant interaction of Comparison Source and Condition ($F_{2,110} = 7.92, p = 0.001$) such that those who participated in the Optimal ordering gave suggestions which were more similar to both our Optimal ordering and the Expert ordering. In summary, the Expert suggested order and the model's Optimal suggested order were closer to recommendations given by participants who had experienced the Optimal ordering (see Figure~\ref{fig:suggested_ordering_distance}).

\section{Conclusion and further work} 
 
We presented a framework for reasoning about the impact of user authentication on the overall usability of a workflow. Our framework is the first to highlight the importance of the fit between a particular user authentication method and the context in which it is performed.  Specifically, we draw on results from cognitive psychology to quantify the impact of switching between tasks that draw on different cognitive resources and use different modalities. 
 
This is a new, disruptive approach to evaluating usability of security solutions, and even systems usability in general. We are sharing this powerful core idea with the community in its preliminary form but we envisage further work in several directions, both on our proof-of-concept implementation of the solver and on the framework itself. We need to develop reliable input tools, such as worksheets and flowcharts, to allow independent designers to perform consistent assignment of numerical values to the features of their tasks. More fundamentally, we would like to develop a ``unit'' (not necessarily just elapsed time; maybe other factors like stress and annoyance might come into it) to measure the usability cost, and a disciplined and justifiable method for expressing in this same unit both the cost of a task and the additional cost of a transition. This will allow the CSP solver to add those sub-costs to compute a globally optimal solution. These additional steps go hand in hand with further user studies and validation of the modelling approach. But the general principles and methods that underlie our framework are already useful and applicable today. 
 
Our framework targets two audiences: designers of secure systems and designers of new authentication schemes. System designers can use the framework as scaffolding that supports the overall design process. This scaffolding encourages the designer to think about how their use of authentication is likely to impact on their users and ultimately on the success of the system.  Similarly, security researchers developing new authentication primitives can use the framework to reason about their solution within a realistic context of use. 

Importantly, the theoretical model output was further validated with a user study. Participants performed better in the optimal ordering, and were more satisfied by the optimally ordered interface. The model's optimal ordering was more similar to the suggested orderings of professional designers, and participants who experienced in the optimal ordering were more likely to further prefer and recommend such an ordering. In this way, we were able to validate the predictions of the theoretical model.
 
The consolidation of results from cognitive psychology on the effects of task switching, and the presentation of these results in a format directly usable by security professionals is perhaps the most useful contribution of our work. 
 
\section{Related work} 
 
 
Sasse \textit{et al.}\ \cite{sasse2014great,steves2014report} present their findings of a 2-part study into the impact of authentication on the productivity of employees in a US governmental organisation. They conclude that the overall burden of user authentication includes a disruption to the user's primary task (that is, what they are actually trying to achieve). Disruptions resulting from user authentication damage productivity and result in significant frustrations. Furthermore the authors found that \textit{avoidance}---not logging into services or using them less frequently---was an increasingly common coping strategy when the burden of authentication was felt to be too great. 
 
While Shay \textit{et al.}\ \cite{shay2014can} have attempted to boost security by pushing the limits of user workload, there is a call for designers to consider the impacts of effortful authentication mechanisms on the user. Employees reported to Inglesant and Sasse \cite{ inglesant2010true} that they'd resort to insecure workarounds in response to increasingly stringent password policies. This “friction” \cite{beautement2009compliance} between the tasks has been shown to moderate individual compliance. 
 
Building on these observations, our work is the first attempt to develop a model of such costs. The ultimate goal of this model is to empower system designers to reason about such effects before deployment. 
 
Prior work has demonstrated the usefulness of modelling subtask arrangement to find optimal orderings. Crampton \cite{   crampton2005reference} arranges security-related subtasks to find orderings that satisfy entailment, cardinality, and role-based constraints. Zhang \textit{et al.}\ \cite{ zhang2012reducing} use an optimization procedure to minimize mouse clicks in a computerized task workflow. Our methodology uses similar techniques to consider a finer grained user-centric cost model to optimize the handing off of cognitive mechanisms throughout a task. 
 
Constraint Satisfaction Problems (CSP) have long found application in decision supports systems. Scheduling---determining the optimum allocation of shared resources to competing activities---is a well-known NP-complete Constraint Satisfaction Problem (CSP) \cite{johnson1982np}. 
 
Cohen \textit{et al.}\ \cite{DBLP:journals/jair/CohenCGGJ14} apply techniques from CSP to the \textit{Workflow Satisfiability Problem   (WSP)}---that is, deciding whether a plan exists for assinging task to authorized users in a given business process. Our work draws inspiration from their use of CSP. However, in our framework we are concerned with an optimization problem. 
 
\ifCLASSOPTIONcompsoc 
  \section*{Acknowledgments} 
\else 
  \section*{Acknowledgment} 
\fi

The Cambridge authors are grateful to the European Research Council for funding this research through grant StG 307224 (Pico). The UCL authors are grateful to the Engineering and Physical Sciences Research Council for funding this research through grant \#EP/K033476/1. We are also grateful to Max Spencer and Jeunese Payne for useful discussions, and to Ben Wong for scheduling and running participants. The authors also thank Graham Johnson for assistance in reaching out to professional designers.



\bibliographystyle{IEEEtranS}
\bibliography{2016-SasseETAL-context}
%




\end{document}